\documentclass[sigconf]{acmart}
\pdfoutput=1

\acmConference[]
\acmYear{2018}

\setcopyright{none}

\renewcommand\footnotetextcopyrightpermission[1]{} 
\def\withComments{0} 

\usepackage{amsmath,amsthm,amsfonts,amssymb}
\usepackage{url}
\usepackage{courier}
\usepackage{ifpdf}
\usepackage[normalem]{ulem}
\usepackage{color}
\usepackage{subfigure}
\usepackage{multirow}
\usepackage[ruled,vlined]{algorithm2e}
\usepackage{tabulary}
\newcolumntype{K}[1]{>{\centering\arraybackslash}p{#1}}
\newcolumntype{L}[1]{>{\raggedright\arraybackslash}p{#1}}
\usepackage{graphicx}

\ifodd\withComments
\else
\newcommand{\Gidon}[1]{}
\newcommand{\danny}[1]{}
\fi

\newcommand{\remove}[1]{}

\begin{document}
\title{Securing the Storage Data Path with SGX Enclaves}

\author{Danny Harnik \hspace{0.07in} Eliad Tsfadia \hspace{0.07in} Doron Chen \hspace{0.07in} Ronen Kat}
\affiliation{%
  \institution{IBM Research -- Haifa}
}
\email{ {dannyh, eliadt, doronc, ronenkat}@il.ibm.com}

\renewcommand{\shortauthors}{Harnik et al.}

\begin{abstract}
We explore the use of SGX enclaves as a means to improve the security of handling keys and data in storage systems.
We study two main configurations for SGX computations, as they apply to performing data-at-rest encryption in a storage system. The first configuration aims to protect the encryption keys used in the encryption process. The second configuration aims to protect both the encryption keys and the data, thus providing end-to-end security of the entire data path.

Our main contribution is an evaluation of the viability of SGX for data-at-rest encryption from a performance perspective and an understanding of the details that go into using enclaves in a performance sensitive environment. Our tests paint a complex picture: On the one hand SGX can indeed achieve high encryption and decryption throughput, comparable to running without SGX. On the other hand, there are many subtleties to achieving such performance and careful design choices and testing are required.

\end{abstract}
%
%

\maketitle

\section{Introduction}
Intel $^{\circledR}$ SGX enclaves\cite{SGX_Web} provide hardware enforced confidentially and integrity guarantees for running computations.  This is mainly achieved by encrypting all information as it leaves the CPU, effectively shielding data in the memory from external observers.
There are numerous use cases for SGX that range from running analytics over encrypted data (e.g. \cite{VC3, OSF+16ML, Princess17, Opaque17}) to improving confidentiality in distributed infrastructures (e.g. Tor, Blockchain, Zookeeper) to running entire applications and ecosystems inside enclaves (\cite{Haven14, Scone16, Graphene17}). While some of these applications include large amounts of code and functionality inside enclaves, the original concept of SGX was created with less ambitious use cases in mind, in which minimal parts of the computation (and the code associated with it) are placed into enclaves. For example, Intel's initial papers \cite{HMR13SGX} describe password protection and DRM use cases.
In this work we study the performance aspects of using SGX for security hardening in a use case in which only a small part of the application needs to reside in an enclave. There are numerous applications and uses-cases where only a limited part of the application needs to read and update the data, and the majority of the application deals with aspects such as storing the data, routing it, or making sure it is highly available or durable. In such use cases we can identify a small part of the application that implements the functionality that works on the data, and use enclaves as secure and tamper-proof silos for performing these operations. Examples of operations could be data transformations, local tests on data, and cryptographic functions on data blocks.
Specifically, the main use case that we study in this paper is that of securing the data path and hardening the security of data-at-rest encryption in storage systems such as block storage systems, file systems or object storage. In these systems the data is mostly considered ``payload'' and the operations on the actual data are very limited, as the goal of the storage system is to make the data accessible, highly available, durable and resilient, requirements which have very little to do with the actual content of the data. Therefore implementing encryption in SGX enclaves is a good match for the use-case we choose. In addition, we especially focus on performance oriented systems in which the throughput of enclave operations is expected to be high, therefore it is important to understand the overhead of running functions in SGX enclaves.
One would expect some overhead due to the added encryption and decryption complexity. In addition, extra security measures such as integrity tests and memory usage limitations can also affect performance. We present a set of micro-benchmarks that
shed some light on the viability of using enclaves vs.\ running outside enclaves.
Our use case avoids workloads that are documented to be bad for SGX performance (see \cite{Scone16, SGXWhitePaper18}. These include workloads that require frequent small random access operations or workloads that require a very large amount of data to be in encrypted memory simultaneously.

The main questions that we ask are:
\begin{itemize}
\item Can SGX enclaves achieve high throughput, comparable to running the same operation without enclaves?
\item What subtleties arise from using SGX for this use-case and how much development work is required in order to get acceptable performance?
\end{itemize}

Our results paint a rather complex picture. On the one hand, we see cases in which the answer to the first question is positive -- running in an SGX enclave can indeed achieve very high throughput (as much as $90-99\%$ of running without an enclave). On the other hand, such performance does not come easily. We observe that there are many subtleties that affect enclave execution, and note that simply running what seems to be a perfectly good implementation can at times result in very low performance. Key factors such as block size, multi-threading and the exact library involved are key to achieving acceptable performance.


\section{Background and Related Work}\label{sec:back}
There are several main components to SGX that allow it to provide a higher level of security than running on ``untrusted" CPU. A central mechanism is the Memory Encryption Engine (MEE) which performs real time encryption of all communication between the CPU and the memory. SGX enclaves have the capability to access the entire main memory of the machine. However, the MEE is only
invoked on a special designated area of the memory called the Enclave Page Cache (EPC).
This memory area is a very small space out of the entire memory, and in most existing systems is limited to 128MB (or even 96MB after reducing space used for managing the enclaves).

SGX performance overheads can result from a number of reasons \cite{SGXWhitePaper18}: first is the actual overhead of accessing the encrypted memory via the MEE. A second reason is the overhead associated with entering and exiting an enclave. In order to ensure the isolation of these processes, enclaves are only invoked via a special interface called ECALLs (defined in an "edl" file). The ECALLs are known to have a performance impact due to the CPU's context switches. Several works including \cite{Eleos17, Hotcalls17} work to minimize this overhead by avoiding ECALLs as much as possible.
There is also the limitation of the EPC size, and when operating with memory that exceeds the EPC size there is a need for paging of EPC pages to regular memory, introducing significant additional latency. This operation requires both data migration as well as encryption of data before it lands in regular memory (and decryption when pulled back in). Some works manage to improve the efficiency of this process \cite{Scone16,Eleos17} but cannot remove the overhead altogether. Another source of overhead is in the handling of cache misses and specifically the mechanisms that prevent replay attacks on enclaves \cite{Gueron16}. As a result, ``cache non-friendly" workloads are expected to suffer more inside enclaves than in regular operation.
In our work we manage to observe all of these phenomena, and attempt to actually quantify their effects for our use-case.

Additional related works include various implementations of systems that run overall performance tests of their implementations. Examples of such works include \cite{Opaque17, LightBox17, WWBWWW17}. Our work attempts to quantify the performance effect of the storage encryption use-case, and understand the subtleties involved that may affect a wide array of other use cases.

\section{ Storage Systems Encryption }\label{sec:at-rest}

Our study is motivated by the world of storage systems in which data-at-rest encryption
is now a prevalent practice. This practice involves data arriving at the storage system, either in clear text or encrypted in transit (e.g., IPSEC or HTTPS), and then encrypted before being persisted to disk. Its purpose is to ensure that all persistent data is always encrypted, so that loss of hardware, either due to hardware failure or due to malicious behavior, does not compromise the data.

One common approach to handling the encryption/decryption is the use of Self Encrypting Drives (SEDs). In such drives, a special cryptographic processor is built into the hard drive's circuit and performs all encryption and decryption operations (typically an AES encryptor). The encryption keys are communicated over a secure channel into the drives rendering them unreachable without tampering with the SED hardware protection. However, in the absence of SEDs, many systems rely on software based encryption and decryption (e.g. \cite{DMCrypt, GPFSEncryption, ZFSEncryption}). In order to perform software encryption and decryption using the CPU, the encryption keys must reside in clear text in memory, presenting a significant security risk. Sensitive data keys are vulnerable to either privileged users or to memory sniffing techniques.

\begin{figure}[ht!]
\centering
\vspace{-0.1in}
\includegraphics[trim=0 0 500 60,clip, width=0.4\textwidth]{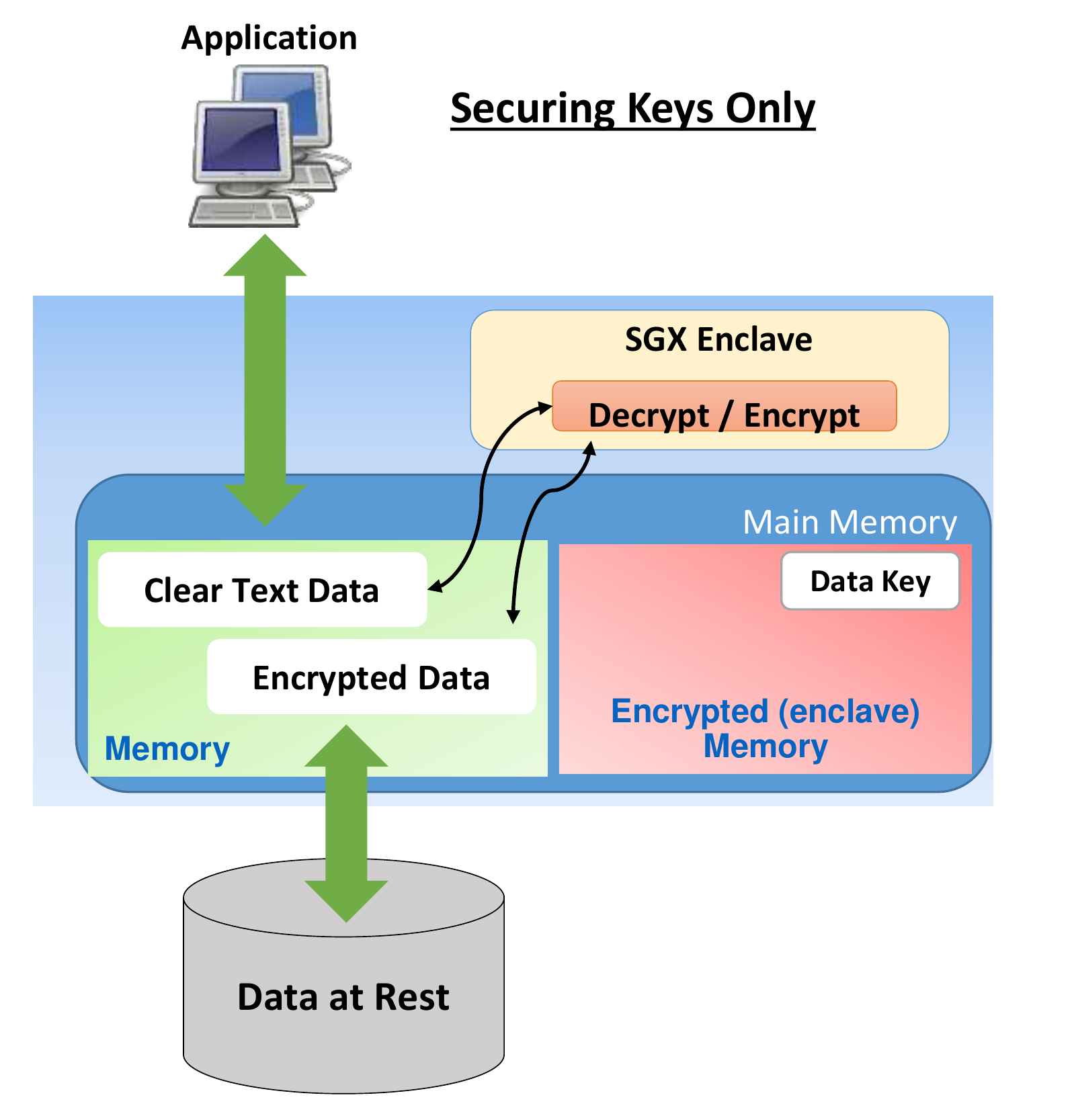}
\vspace{-0.2in}
\caption{
When only keys are secured by the enclave, both encrypted and decrypted buffers are in the general memory. }\label{fig:keys}
\end{figure}

\subsection{Protecting the Keys Using SGX}\label{sec:keys}
By placing the software encryption process and key handling inside enclaves, one achieves a much stronger protection: keys are never in the clear -- they are communicated over a secure channel into the enclave, and placed in encrypted memory in which all encryption and decryption is performed.
Note that in this solution, the protection is of the {\em data encryption keys} rather than of the data itself. For systems in which data arrives in the clear (rather than over a secure channel) this makes perfect sense. Data is only protected at rest, but the data keys (which are inherently more sensitive) are encrypted at all times. This is mostly the case today due to performance limitations of the storage clients.

When working in this setting, it should be noted that data does not need to reside in the enclave's encrypted memory. Enclaves are allowed to access the general memory, and so the data buffers (both encrypted and cleartext) can reside outside of the enclaves memory. An encryption request to the enclave is performed via an ECALL that gets pointers to two buffers in the general (non-encrypted) memory. This is an important subtlety which has a performance impact which will be discussed and demonstrated in Section~\ref{sec:aes}. This solution is depicted in Figures~\ref{fig:keys} and provides equivalent security guarantees as SEDs provides for the data keys.
\begin{figure}[ht!]
\centering
\vspace{-0.1in}
\includegraphics[trim=0 0 500 50,clip, width=0.4\textwidth]{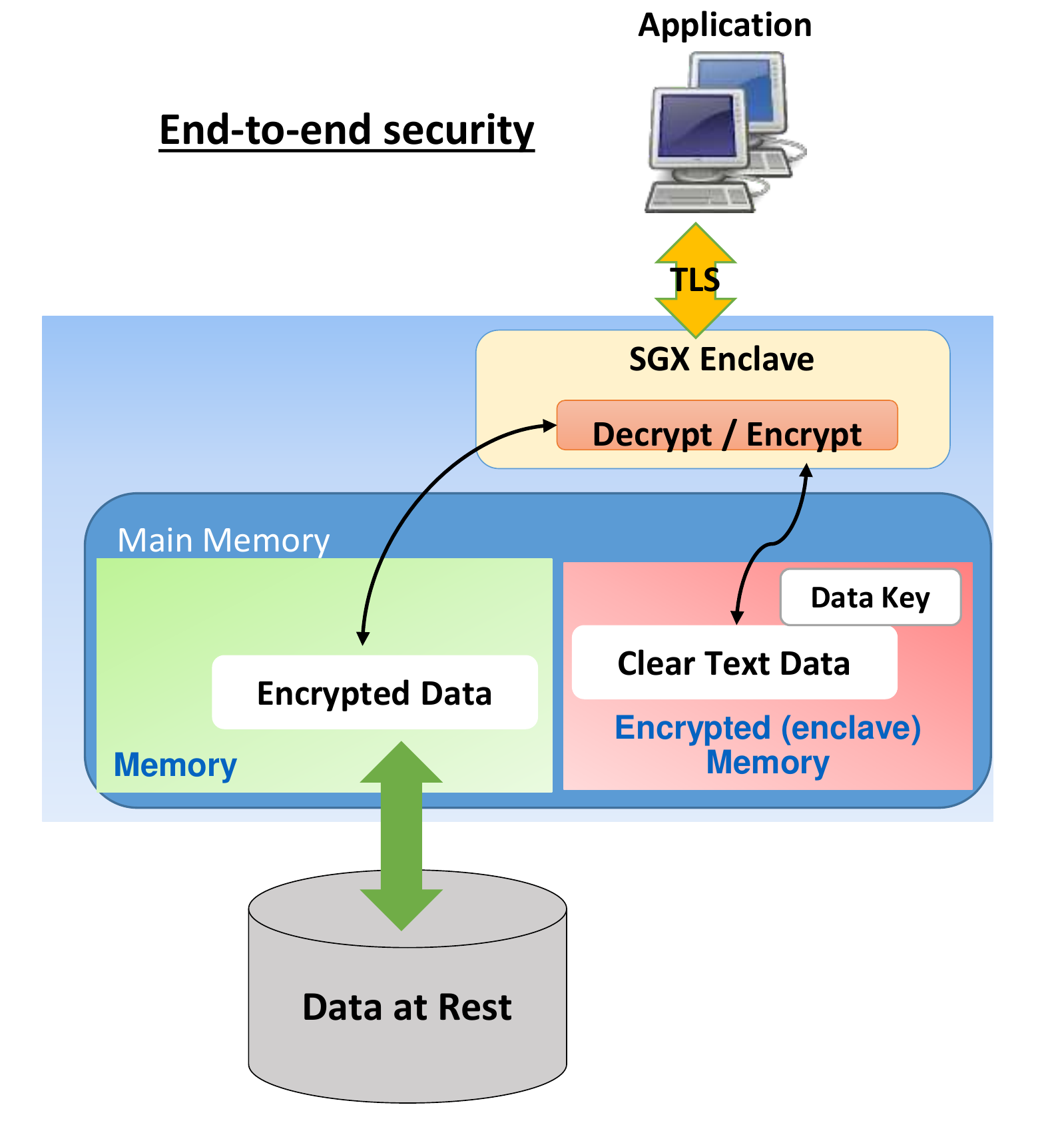}
\vspace{-0.2in}
\caption{
In the end-to-end variant, the clear text data must reside inside the enclave memory. In reality, there may be another temporary encrypted buffer to hold the TLS encrypted data. This buffer is encrypted and hence can reside in the general memory.}\label{fig:e2e}
\end{figure}

\subsection{End-to-End Data protection}\label{sec:e2e}
Data-at-rest security as a sole encryption protection is quite a common configuration in many enterprise settings. However, this is changing with the growing awareness to threats, the movement to the cloud and the improvement of encryption capabilities. Data-in-transit encryption is now required in many scenarios, typically via one of the prevalent standards such as HTTPS, SSL, TLS, FTPS, etc. In this mode of operation, a secure channel is negotiated between a client and a server through which data is encrypted for the duration of the transfer, but decrypted upon arrival. While the data is now protected traveling through the network until arriving to the storage system and when residing on disk, this does leave the data in the clear while in memory until it is re-encrypted for data-at-rest protection. This is where SGX enclaves can close the gap and create a full end-to-end security of data --- data is received and decrypted only inside the SGX enclave and encrypted for at-rest using keys stored in the enclave. Therefore, from the time data leaves the client to the storage until it is read again, it is kept encrypted in all stages. Together with the advantages of key protection described in the previous section, using enclaves for end-to-end security is a very appealing notion.\footnote{It should be noted that an alternative to achieving end-to-end security is to simply encrypt data at the client side (outside the storage), and never decrypt it in the storage. However, in the case of multiple clients (i.e., machines) accessing the same data, this moves key management complexity to the client side (since all clients are required to encrypt and decrypt using the same keys) making this alternative unpopular.}

We note that such a strong notion of security cannot be achieved using SEDs, since the termination of the secure channel cannot be done at the disk level and therefore data is decrypted in memory before being persisted. In order for enclaves to close the gap, it is required that both the establishment of the secure (e.g. the TLS handshake) and the termination of the protocol must happen inside an enclave. We note that several works and open source projects deal with how to do this inside an enclave (e.g. \cite{Towncrier16,aublin2017talos,wolfSSL-SGX,sgx-ra-tls,abs-1801-05863}). Figure~\ref{fig:e2e} describes the flow of this solution.

\section{Evaluation}\label{sec:eval}

Our benchmarks were tested in the following setup: We used two testing setups, a Lenovo server with 4-core Intel $^{\circledR}$  Xeon $^{\circledR}$ Processor E3-1270 v5 CPU with 3.60GHz, 16GBRAM, and a Lenovo laptop with a 4-core Intel Core $^{\circledR}$ i7-6820HQ CPU with 2.70GHz, using 32GB RAM. The software ran on an Ubuntu 16.04 OS and the libraries that we used were the Intel SGX Linux 2.0 \cite{sgx-sdk} and SGXSSL \cite{intel-sgx-ssl} (taken from the intel-sgx-ssl git repository on Nov 29th 2017).

Our goal was to evaluate SGX with respect to AES encryption, which is required for our Section~\ref{sec:at-rest} use case. However, AES encryption is a complex function that requires specialized CPU commands and its performance can vary significantly with different implementations (as will be seen in Section~\ref{sec:aes}). In particular, common AES implementations do not seamlessly compile on SGX and their libraries need to be slightly modified in order to run inside SGX enclaves. Instead of jumping directly to AES, we start by evaluating a toy example first -- we take a very simple function with the only requirement that it touches the entire input buffer. Specifically, we tested the overhead of finding the maximum 4-byte integer of a given byte array (i.e., we treat an array of N bytes as an array of N/4 integers).
Our objective is to identify and evaluate the main performance obstacles of running inside an SGX enclave.

\subsection{Testing a Toy Example}\label{sec:max}

We implemented four variations of the function \verb"find_max":
\begin{enumerate}
\item	A regular function running in the {\bf untrusted} area as one would run it without enclaves.
\item	{\bf Copy and compute version} - the array is copied into the enclave's encrypted memory and then iterated over. This is implemented using an ECALL, in which the input array is declared with the "in" option in the edl file.
\item	{\bf Compute on encrypted memory} - in this option we find the maximum on an array that resides in the enclave's encrypted memory. The array is prepared before the ECALL (by a previous ECALL). This is similar to option 2, but does not include the initial buffer copying operation in the measurements. This usages matches the end-to-end encryption setting described in Section~\ref{sec:e2e}.
\item	{\bf Compute on cleartext memory} - in this option the ECALL finds the maximum of a given external input array without copying it into the enclave's memory (namely, accessing only clear-text memory). This is achieved by declaring the array with the ''\verb"user_check"" option in the edl file. This usage corresponds to the protecting of keys only mode described in Section~\ref{sec:keys}.
\end{enumerate}

Of the 3 enclave variants (variants 2, 3, and 4), option 4 should be the fastest, as it does not require the array to be decrypted by the SGX's Memory Encryption Engine (MEE). Options 2 and 3 do require the MEE decryption in order to perform the computations (while option 2 also involves MEE encryption as well as decryption).

\begin{figure}[ht!]
\centering
\includegraphics[width=0.5\textwidth]{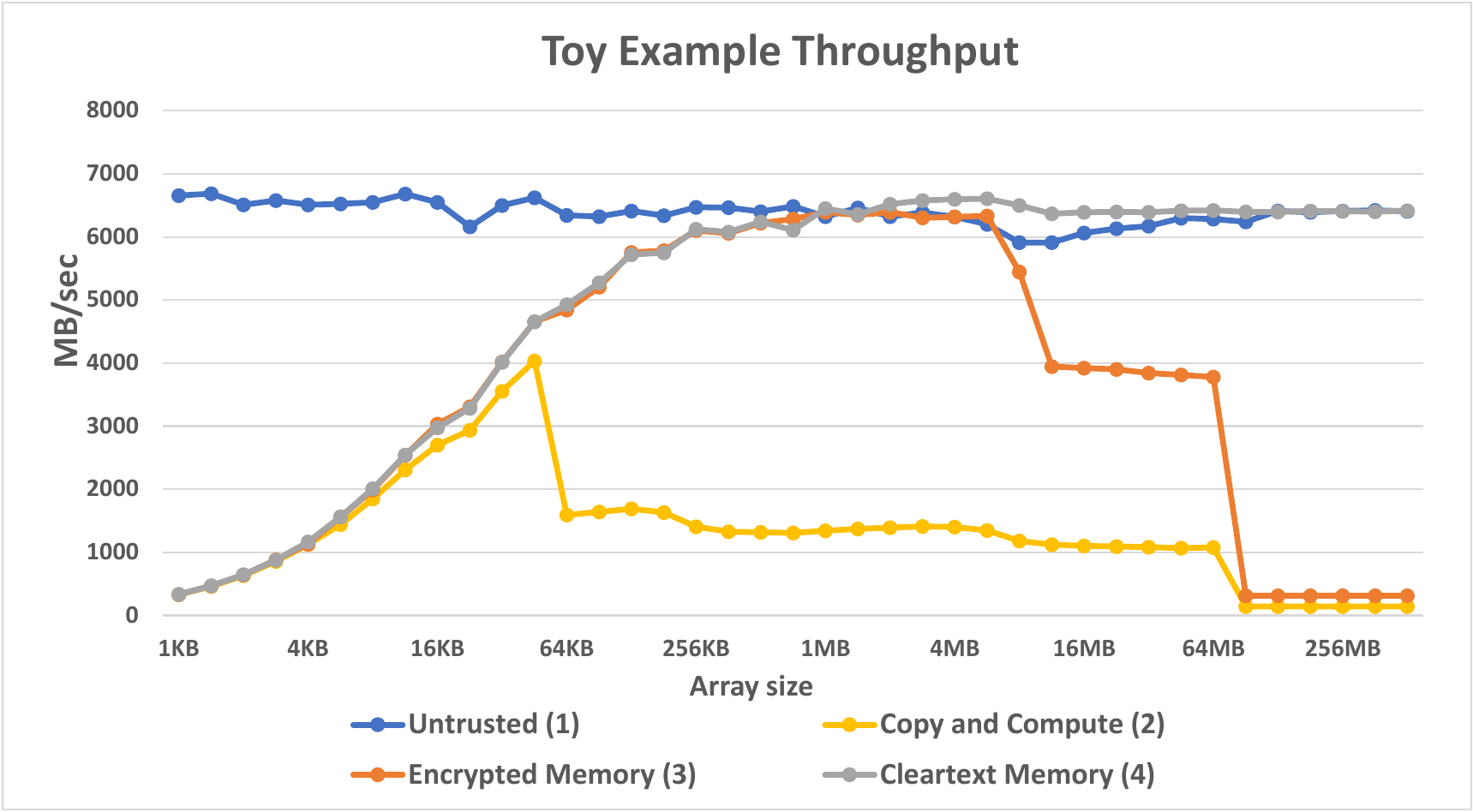}
\vspace{-0.25in}
\caption{The processing throughput of the toy example function ``find max".
}\label{fig:findMax}
\end{figure}

  By evaluating the performance of these three options we gain a pretty good understanding of the expected overhead of ECALL context-switches, the overhead of the MEE operations, and what is the overhead of copying data using the "in" (or "out") declarations in the edl. These observations should hold for other computations other than just the "\verb"find_max"" function.
For evaluating the performance of each option, we compared the throughput of the calls using various array sizes. The results in Figure~\ref{fig:findMax} show the throughput (MBs processed per second) as a function of the array size. These numbers are a single thread test (multiple threads will be discussed in Section~\ref{sec:threads}).
The first observation is that for small arrays, there is a huge overhead in running a function in an enclave; this is likely caused by the context switches overhead of entering and exiting the enclave. This overhead becomes negligible for larger buffers and the gap between the untrusted and the two faster trusted versions (versions 3 and 4) is mostly closed with arrays of size larger than 256KB.

There are three other interesting phenomena: First, for arrays of size larger than 8MB, we see a throughput degradation in the ECALL that runs on enclave encrypted memory. This is related to additional L3 cache misses when reading an enclave's local array (in our setup, L3 cache is of size 8MB). Second, for large array sizes that are greater than the size of the enclave page cache (EPC) (96MB), there is a dramatic drop off. This is the expected drop caused since the EPC can no longer hold the entire buffer in encrypted memory and a paging mechanism is invoked to page this data in and out of regular memory (encrypting and decrypting on the way). Finally, the ECALL that uses the "in" option has a significant slowdown for arrays of size larger than 64KB. Our limited investigation indicates that this may be caused by a slowdown when calling the function memcpy() inside an enclave with buffers of size larger than 64KB.

\section{Testing AES-GCM Encryption}\label{sec:aes}
We turn to test the overhead of encrypting and decrypting messages. In particular, we focused on testing AES128-GCM encryption.
The basic methodology that we used was to create a buffer in memory and encrypt it a large number of times using either a regular function or an ECALL. Hence the buffer is ``hot" in memory, i.e., cached in the CPU, which should be the case in a storage use-case in which a buffer is encrypted upon its arrival and creation. \footnote{We note that we also ran some tests with ``cold" buffers and these revealed that performance inside enclaves suffers a bigger slowdown than running in regular mode.} Each test is run 30 times and the results presented here are averages. In this section we present the finding for a single threaded execution (multiple thread results are presented in Section~\ref{sec:threads}).
Note that the results for decryption (rather than encryption) were very similar so we only present the encryption numbers here.

 For the ``trusted" variants (inside an enclave), we evaluated two libraries: sgxsdk \cite{sgx-sdk} and sgxssl \cite{intel-sgx-ssl}.
 In the untrusted area we tested openssl \cite{openssl}.

\begin{figure}[ht!]
\centering
\vspace{-0.05in}
\includegraphics[width=0.5\textwidth]{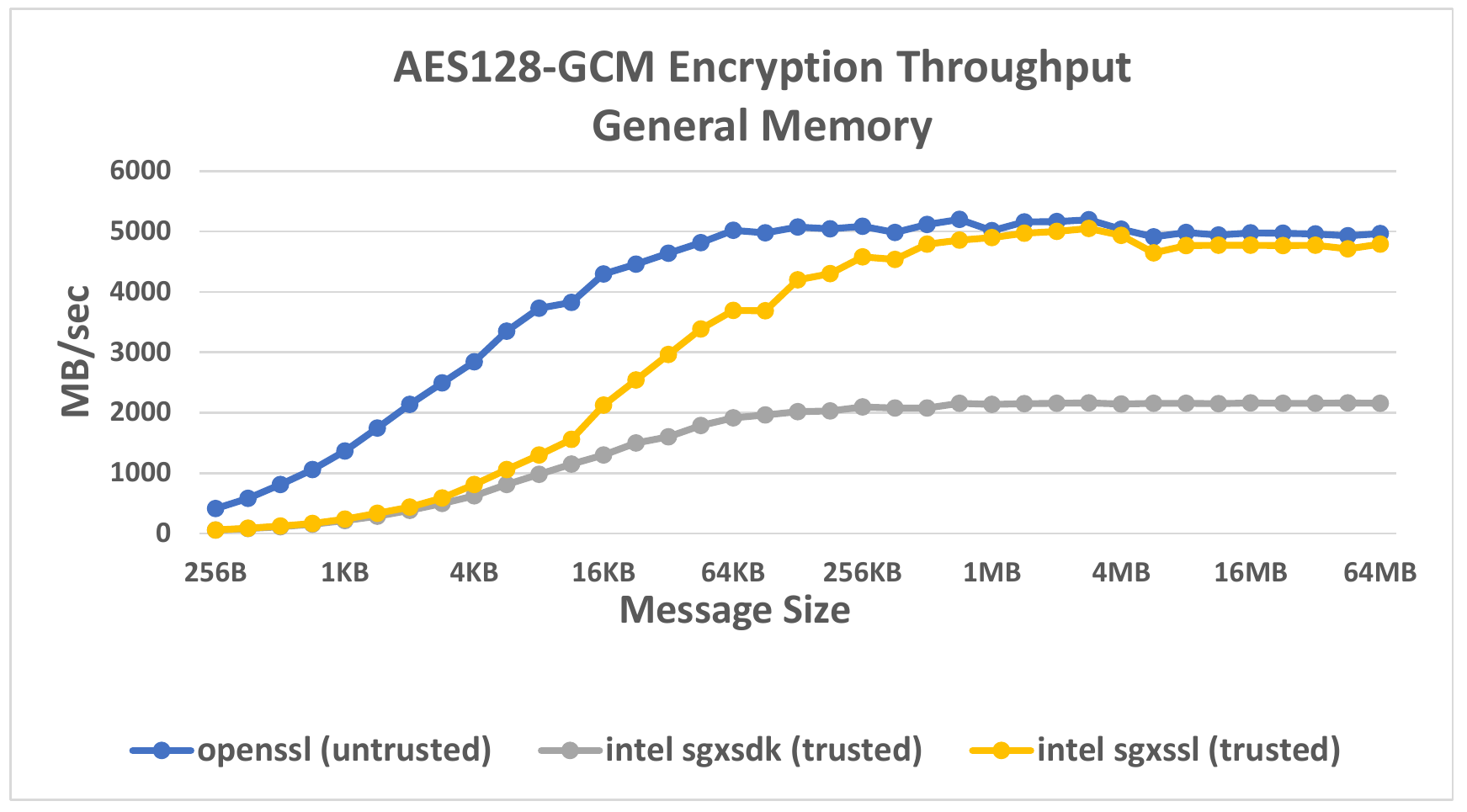}
\vspace{-0.25in}
\caption{Encryption throughput of three code variants for the key-protection scenario in which the buffers are in the general memory.
}\label{fig:enc_non_local}
\end{figure}

We first focus on the key protection  variant of encryption in which the encryption and decryption buffers are in regular memory. This uses an ECALL implementation in which both the input buffer and the output buffer are declared with the "\verb"user_check"" option.
The results, shown in Figure ~\ref{fig:enc_non_local} showed the following behavior:
\begin{itemize}
  \item The sgxsdk version achieves a maximal throughput of about 43\% of the untrusted throughput. This is due to the fact that by default it runs a non-optimized version of Intel's IPP Crypto library which does not use Intel's AES-NI hardware optimizations.
	We were told that by manually compiling and linking the SDK with an optimized binary of the IPP Crypto for SGX, one might achieve the desired acceleration.
	\item The sgxssl test showed an expected behavior: on short messages, we have a significant gap between the trusted and untrusted throughput, which is caused by the context switches overhead. However, for large messages, the trusted version essentially closes the gap with the untrusted library, achieving around 96\% of the throughput of untrusted version (at a rate of about 5GB/sec).\footnote{The extremely high AES encryption that we see can be attributed to optimizations of the Skylake processor, the use of the AES-NI hardware acceleration instruction set, and the fact that our tests encrypt data that probably reside in the cache.}
	\item It should be noted that the initial performance that we saw for the sgxssl implementation was extremely slow and achieved only a 110MB/sec throughput (about 2\% of the untrusted throughput). After sharing this information with Intel the problem was identified and there exists a patch to fix it (see documentation in \cite{intel-sgx-ssl}).
\end{itemize}

\begin{figure}[ht!]
\centering
\includegraphics[width=0.5\textwidth]{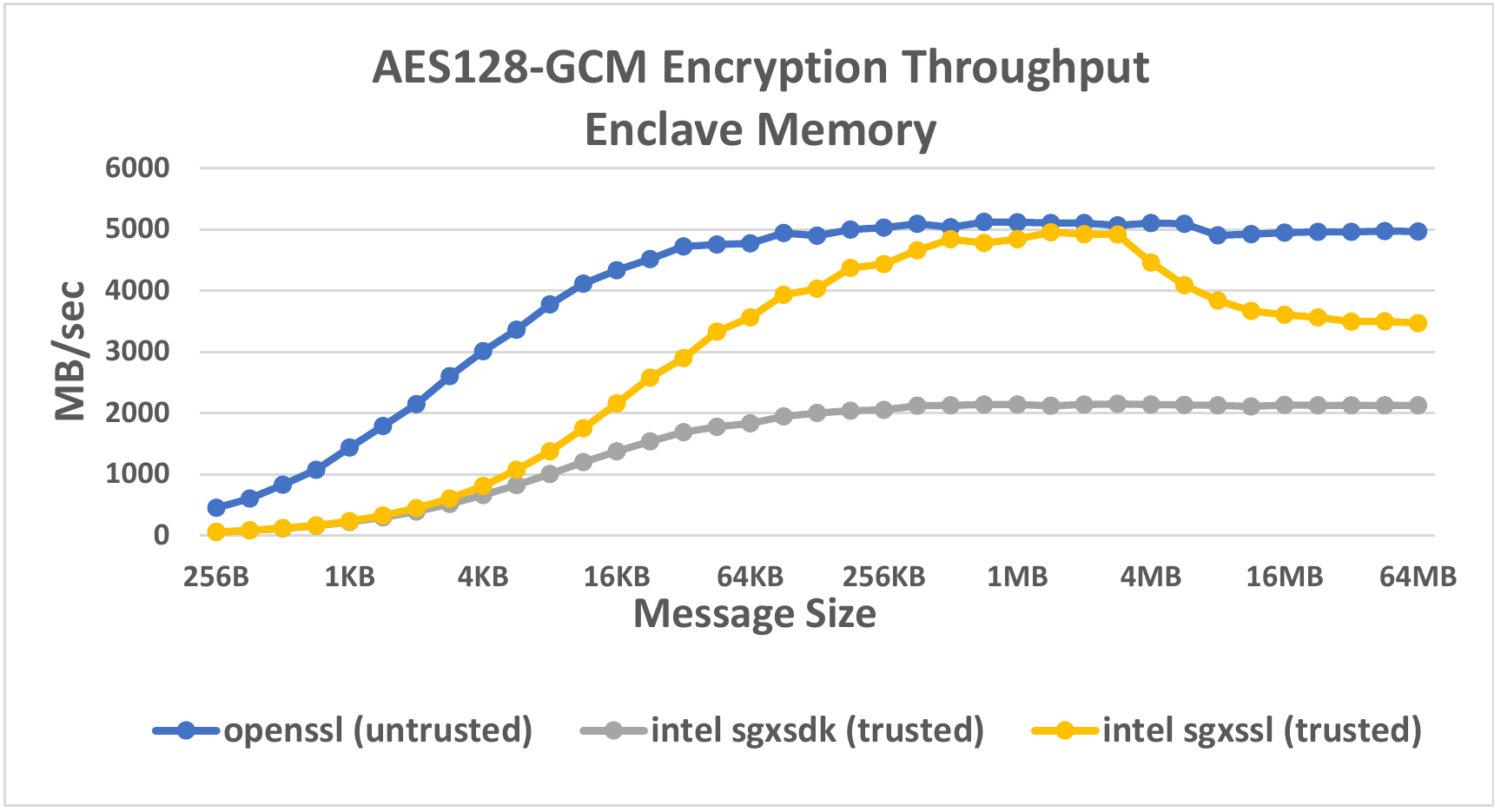}
\vspace{-0.25in}
\caption{Encryption throughput of three code variants for the end-to-end scenario in which the data buffer is in the EPC.
}\label{fig:enc_local}
\end{figure}

We then tested the encryption in the end-to-end use-case, in which the input messages reside in the enclave's encrypted memory rather than clear-text input buffers from the untrusted memory. Note that in this test we did not simulate the arrival of the buffer into enclave memory (which should involve a decryption process), but rather generate this buffer in advance and test the encryption time. The results, in Figure~\ref{fig:enc_local}, are very close to those of the key-protection variant, except for a significant difference for the large buffers.  The sgxssl code suffers from a throughput degradation on messages of size larger than 8MB, which is attributed to cache misses on the L3 cache (a similar degradation to the degradation observed in the \verb"find_max" test). This reduces the maximal throughput of the enclave variant on large buffers to about 70\% of the throughput of untrusted version. Growing the buffer further (above the EPC size threshold of 92GB) results in a massive degradation in trusted code throughput (this is not shown in the figure).

An interesting note is that the degradation due to the L3 cache miss only affects very high performance code and is not seen in ``slower" code running inside enclaves. For example, the sgxsdk version which achieves approximately 40\% of the maximal throughput does not show degradation (this was also the case for SHA256 tests that we ran and achieve a lower throughput altogether).

\subsection{The Effect of Multiple Threads}\label{sec:threads}
The evaluation of a single thread offers only part of the picture and we move to explore the impact of running multiple concurrent threads. Our tests from here on focus just on the fastest  library versions, the ``trusted" (inside an enclave) and ``untrusted".\footnote{In particular we used the encryption library from  the open source project Opaque~\cite{Opaque17} which achieves similar performance to sgxssl. This was instead of using the sgxssl version, which was not fixed yet at the time we ran our tests.} We tried running several configurations: (1) multiple processes; (2) multiple threads with a single enclave; and (3) multiple threads with separate enclaves. We did not see any significant difference between the results of these configurations and we present results for the second configuration of threads using a single enclave instance.
The methodology that we used for testing is to create a number of threads and run repeating encryption tests for all of them on their respective buffers. We run included a warm-up and cool down to ensure that all thread measurements are done with all threads executing simultaneously. This test is aimed at measuring the maximal throughput of a fully utilized system.

\begin{figure}[ht!]
\centering
\includegraphics[width=0.5\textwidth]{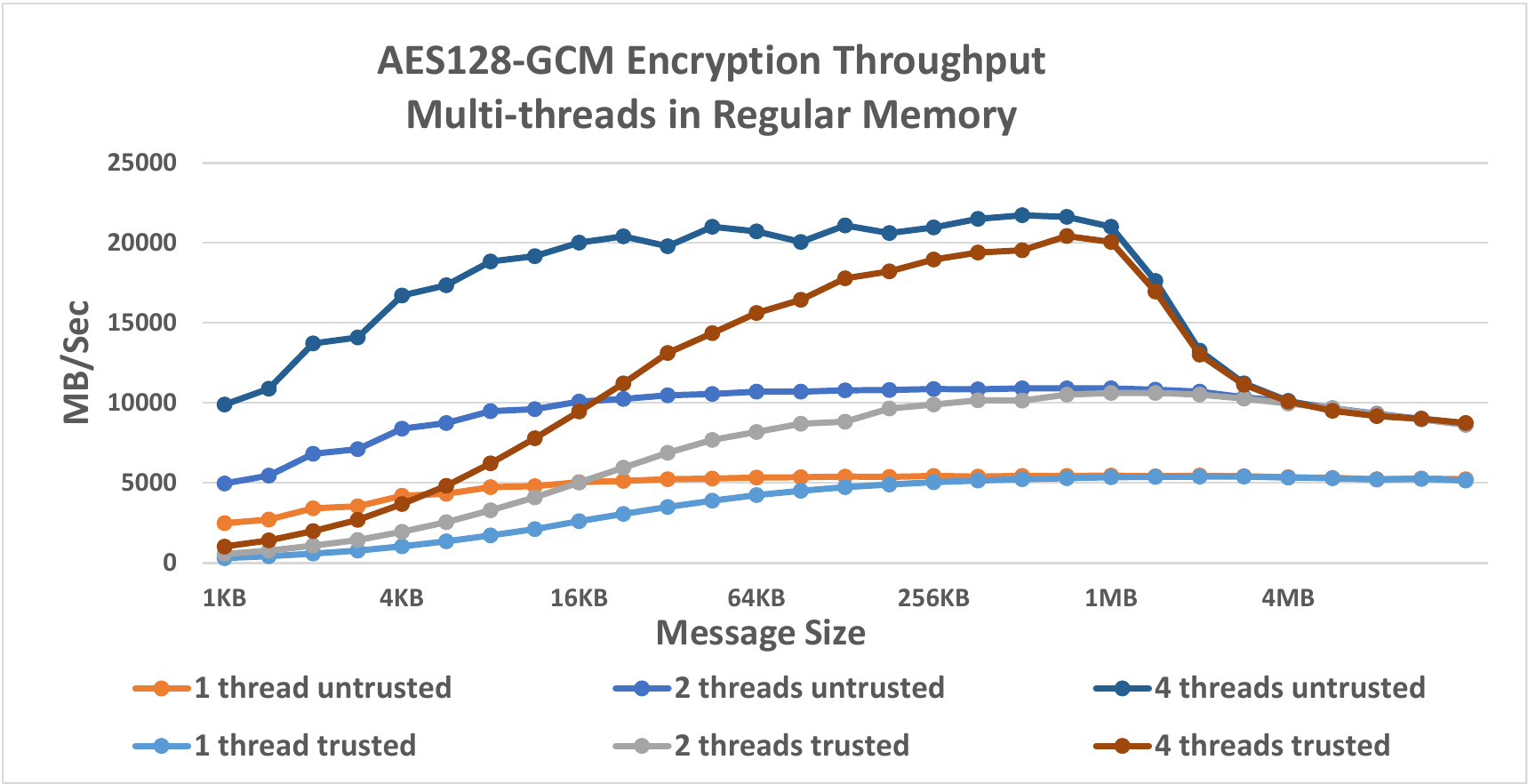}
\vspace{-0.25in}
\caption{Encryption throughput for 1, 2 and 4 threads for the regular memory setting. Cache issues cause the multi-threaded operation to drop for large buffers for trusted and untrusted alike.
}\label{fig:mt_non_local}
\end{figure}

\begin{figure}[ht!]
\centering
\includegraphics[width=0.5\textwidth]{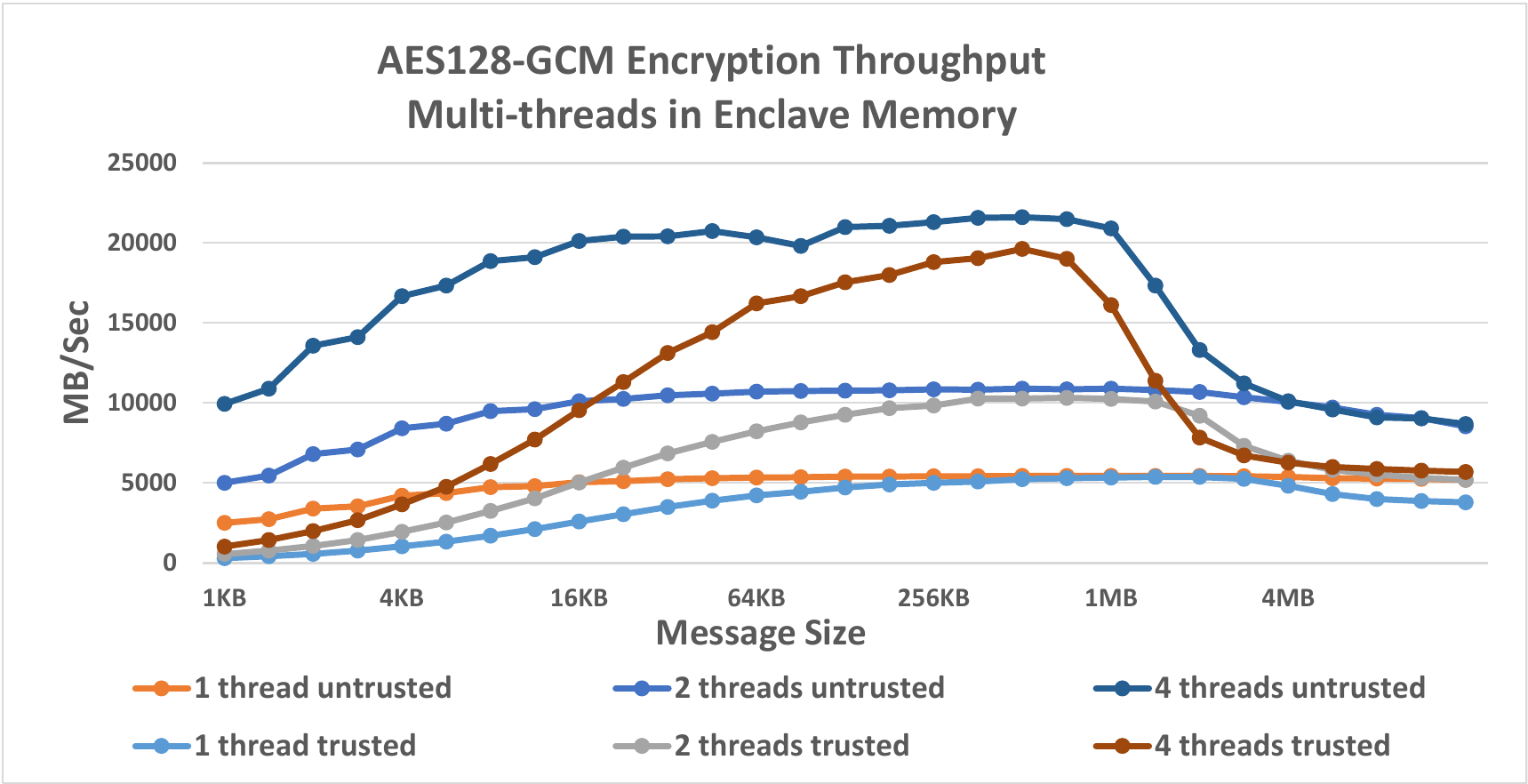}
\vspace{-0.25in}
\caption{Encryption throughput for 1, 2 and 4 threads for the encrypted memory setting. Notice that with 4 threads the trusted version does not reach the untrusted numbers.
}\label{fig:mt_local}
\end{figure}

Results for the key protection use case (regular memory) are in Figure~\ref{fig:mt_non_local} and for the end-to-end use case are in Figure~\ref{fig:mt_local}. The first thing that pops out when looking at the results is the strong effect that the buffer size has on the multi-threading performance regardless of enclaves and SGX. Namely, we see a bit dropoff when running multiple threads once the buffers cross the 1MB threshold. This becomes even more noticeable as the number of threads grows (e.g. for 8 threads the drop off happens once the buffer size exceeds 512KB). The good news is that this effect is similar both inside and outside enclaves. For the regular memory test (Figure~\ref{fig:mt_non_local}), the trusted execution in enclaves reaches very similar results to that outside enclaves. This happens once the buffer is large enough to overcome the costs of the ECALL context switches. In the encrypted memory case (Figure~\ref{fig:mt_local}), the situation changes and running inside enclaves does take a toll. In particular, the trusted execution at the best buffer size loses approximately 10\% of the potential throughput when running with 4 threads.

\begin{figure*}[ht!]
\centering
\vspace{-0.2in}
\hspace{-0.2in}
\includegraphics[width=0.33\textwidth]{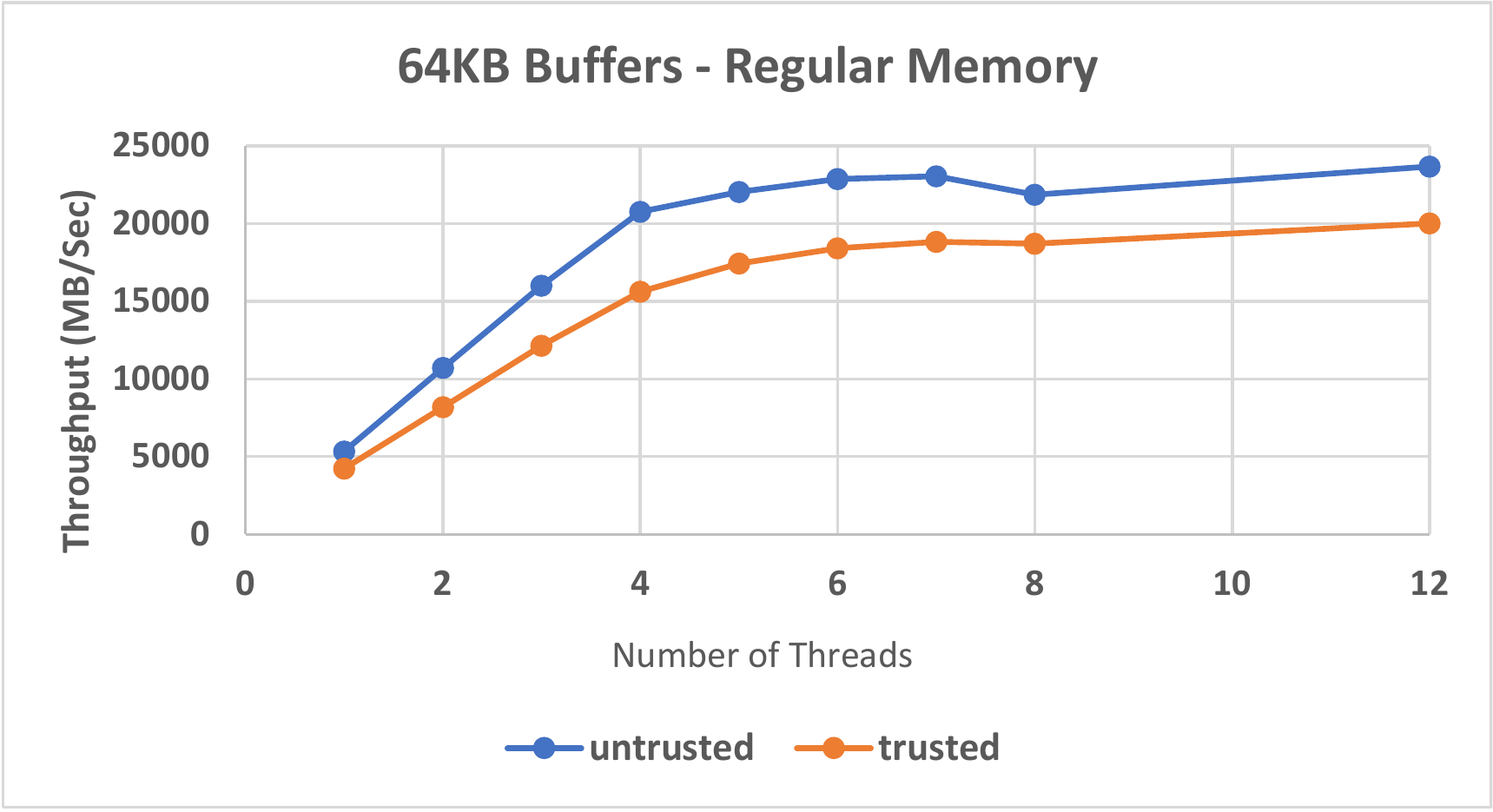}
\includegraphics[width=0.33\textwidth]{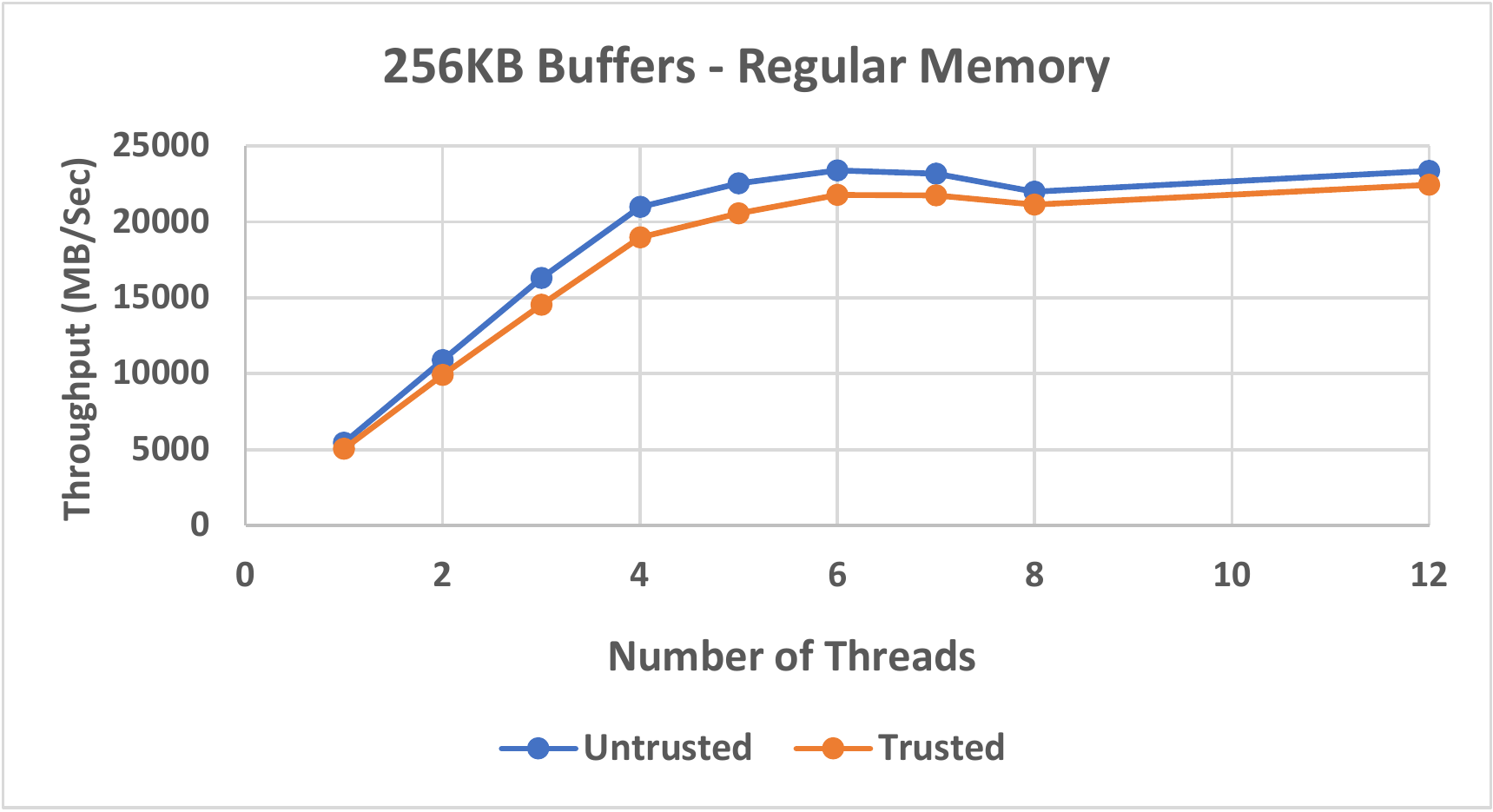}
\includegraphics[width=0.33\textwidth]{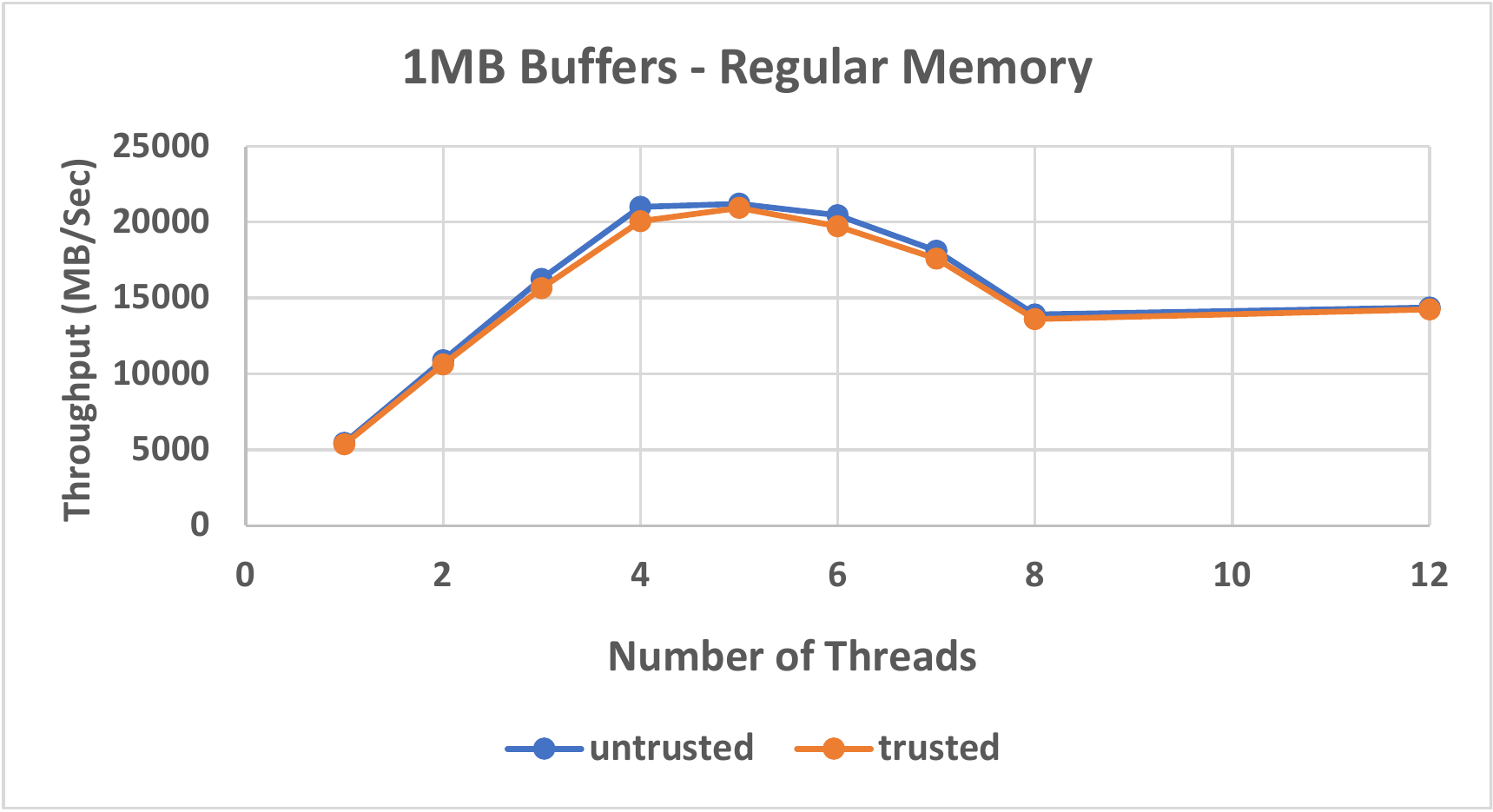}\\
\vspace{-0.2in}
\hspace{-0.2in}
\includegraphics[width=0.33\textwidth]{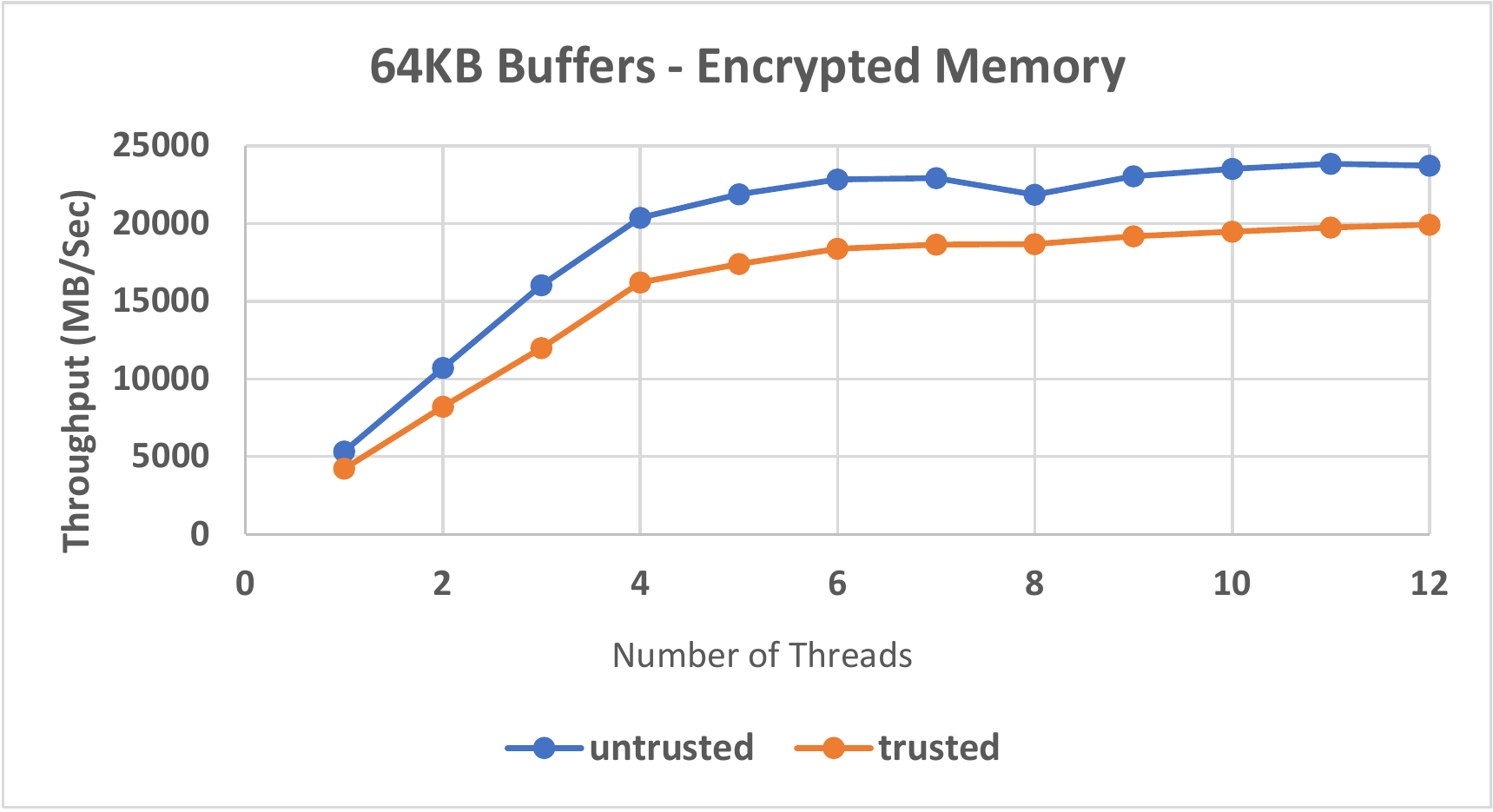}
\includegraphics[width=0.33\textwidth]{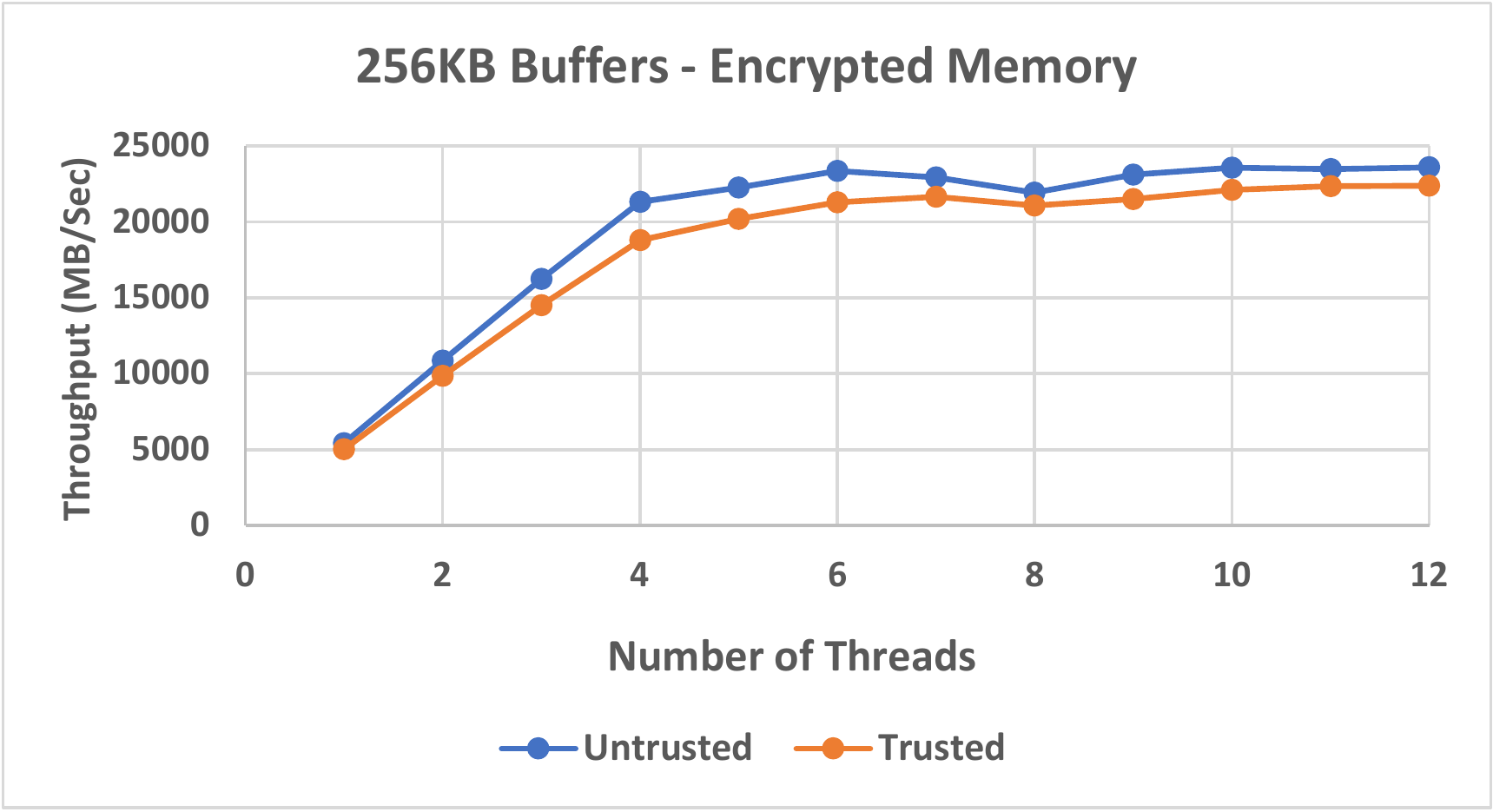}
\includegraphics[width=0.33\textwidth]{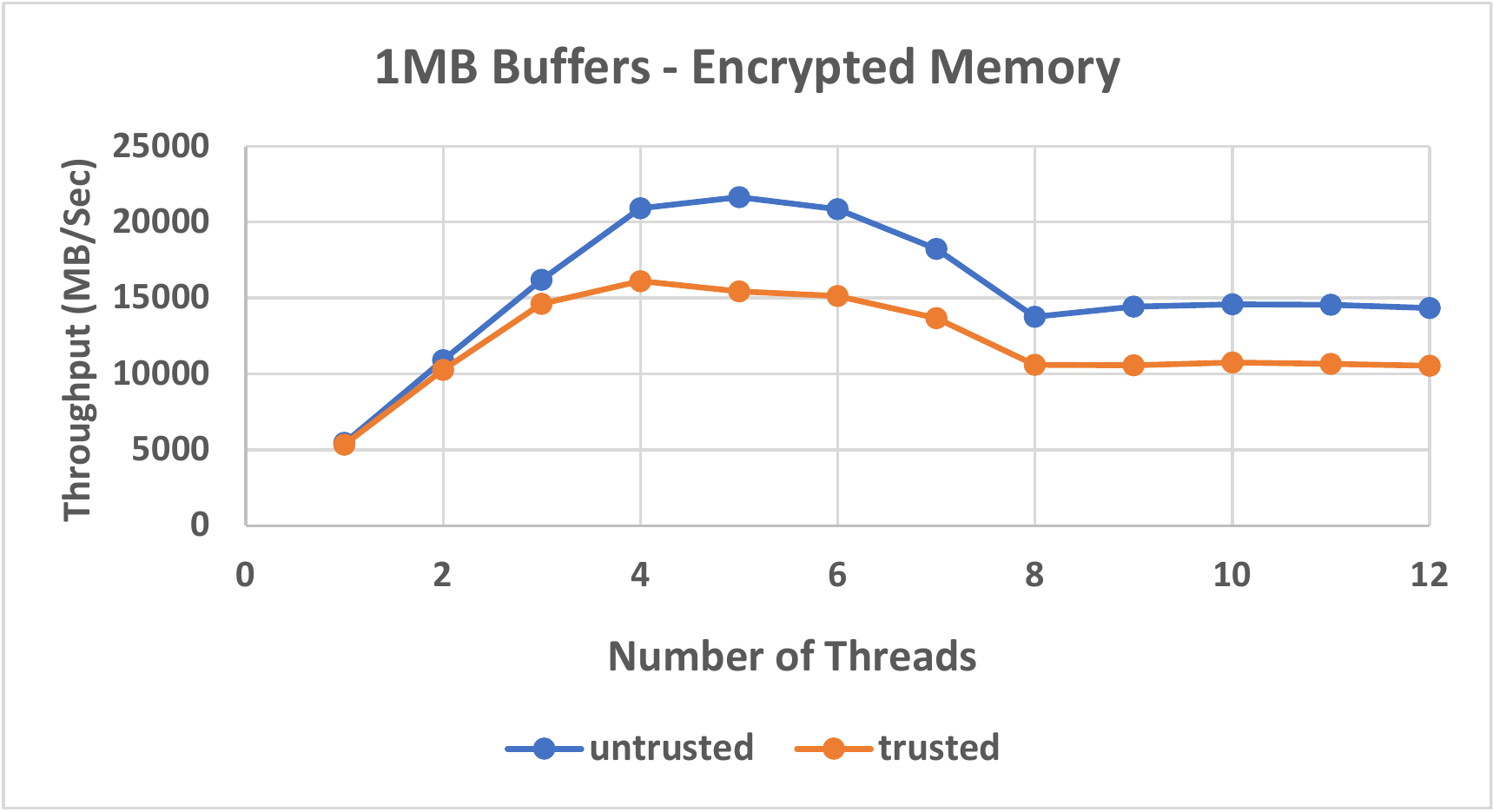}\\

\caption{ The throughput for various buffer sizes as a function of the number of threads. The top row relates to the key protection model and the bottom row to the end-to-end encryption.
  }\label{fig:threads}
\end{figure*}

Figure ~\ref{fig:threads} shows the effect of growing the number of threads beyond 4 on a few select buffer sizes.
The machines on which we tested had 4-cores and 8 thread hardware support. So we expected the tests to achieve up to a factor of 4 on the single thread performance, and were unclear what additional threads would gain. This was very close to what we saw in practice and indeed the performance is near linear for the first 4 threads and then plateaus and even declines. Here too we can see that running on encrypted memory takes a toll and the trusted execution does not catch up to the untrusted run.

\section{Conclusions and Discussion}\label{sec:conclusions}
On the positive side, we see that by choosing correct buffer sizes and correct library configurations, SGX enclaves can achieve very high throughput, on par with code running outside enclaves.

On the other hand, we see that nothing is simple with SGX enclaves. First, using the default crypto libraries provided by Intel $^{\circledR}$, or any library ported for SGX, does not guarantee optimal performance. Second, even when we have a code that achieves an optimal throughput when it accesses a clear-text data, it might not achieve that when it accesses an enclave's local memory. In fact, when running on an enclave's memory, there seems to be a limited sweet spot, where the input should not be too small nor too large.
There are many subtleties here and depending on the number of threads and other work being run in the background, finding the correct configuration can be tricky. In particular, large buffers suffer from the strong effect of cache misses on enclaves, while small buffers suffer from ECALLs. Using mechanisms for avoiding ECALLs (e.g. \cite{Eleos17, Hotcalls17}) may have a positive effect improve the ability to achieve high throughput also when many threads or other work in the system are challenging the L3 cache.
Our strongest impression was that without actual performance testing, it is difficult to predict the performance of computations running inside an SGX enclave.

\bibliographystyle{ACM-Reference-Format}
\bibliography{sgx}


\begin{thebibliography}{26}


\ifx \showCODEN    \undefined \def \showCODEN     #1{\unskip}     \fi
\ifx \showDOI      \undefined \def \showDOI       #1{#1}\fi
\ifx \showISBNx    \undefined \def \showISBNx     #1{\unskip}     \fi
\ifx \showISBNxiii \undefined \def \showISBNxiii  #1{\unskip}     \fi
\ifx \showISSN     \undefined \def \showISSN      #1{\unskip}     \fi
\ifx \showLCCN     \undefined \def \showLCCN      #1{\unskip}     \fi
\ifx \shownote     \undefined \def \shownote      #1{#1}          \fi
\ifx \showarticletitle \undefined \def \showarticletitle #1{#1}   \fi
\ifx \showURL      \undefined \def \showURL       {\relax}        \fi
\providecommand\bibfield[2]{#2}
\providecommand\bibinfo[2]{#2}
\providecommand\natexlab[1]{#1}
\providecommand\showeprint[2][]{arXiv:#2}

\bibitem[\protect\citeauthoryear{??}{DMC}{2017}]%
        {DMCrypt}
 \bibinfo{year}{2017}\natexlab{}.
\newblock \bibinfo{title}{dm-crypt: Linux kernel device-mapper crypto target}.
\newblock
  \bibinfo{howpublished}{\url{https://gitlab.com/cryptsetup/cryptsetup/wikis/DMCrypt}}.
    (\bibinfo{year}{2017}).
\newblock


\bibitem[\protect\citeauthoryear{??}{ZFS}{2017}]%
        {ZFSEncryption}
 \bibinfo{year}{2017}\natexlab{}.
\newblock \bibinfo{title}{Encrypting {ZFS} File Systems}.
\newblock \bibinfo{howpublished}{\url{https://docs.oracle.com/}}.
  (\bibinfo{year}{2017}).
\newblock


\bibitem[\protect\citeauthoryear{??}{GPF}{2017}]%
        {GPFSEncryption}
 \bibinfo{year}{2017}\natexlab{}.
\newblock \bibinfo{title}{{GPFS} Encryption}.
\newblock \bibinfo{howpublished}{\url{https://www.ibm.com/}}.
  (\bibinfo{year}{2017}).
\newblock


\bibitem[\protect\citeauthoryear{??}{sgx}{2017}]%
        {sgx-ra-tls}
 \bibinfo{year}{2017}\natexlab{}.
\newblock \bibinfo{title}{Integrating Intel SGX Remote Attestation into the TLS
  connection setup}.
\newblock
  \bibinfo{howpublished}{\url{https://github.com/cloud-security-research/sgx-ra-tls}}.
    (\bibinfo{year}{2017}).
\newblock


\bibitem[\protect\citeauthoryear{??}{SGX}{2017}]%
        {SGX_Web}
 \bibinfo{year}{2017}\natexlab{}.
\newblock \bibinfo{title}{Intel$^{\circledR}$ Software Guard Extensions
  (Intel$^{\circledR}$ SGX)}.
\newblock \bibinfo{howpublished}{https://software.intel.com/en-us/sgx}.
  (\bibinfo{year}{2017}).
\newblock


\bibitem[\protect\citeauthoryear{??}{sgx}{2017}]%
        {sgx-sdk}
 \bibinfo{year}{2017}\natexlab{}.
\newblock \bibinfo{title}{Intel$^{\circledR}$ Software Guard Extensions
  (Intel$^{\circledR}$ SGX) SDK}.
\newblock
  \bibinfo{howpublished}{\url{https://software.intel.com/en-us/sgx-sdk}}.
  (\bibinfo{year}{2017}).
\newblock


\bibitem[\protect\citeauthoryear{??}{int}{2017}]%
        {intel-sgx-ssl}
 \bibinfo{year}{2017}\natexlab{}.
\newblock \bibinfo{title}{Intel$^{\circledR}$ Software Guard Extensions {SSL}}.
\newblock \bibinfo{howpublished}{\url{https://github.com/intel/intel-sgx-ssl}}.
    (\bibinfo{year}{2017}).
\newblock


\bibitem[\protect\citeauthoryear{??}{ope}{2017}]%
        {openssl}
 \bibinfo{year}{2017}\natexlab{}.
\newblock \bibinfo{title}{OpenSSL, Cryptography and SSL/TLS Toolkit}.
\newblock   (\bibinfo{year}{2017}).
\newblock
\urldef\tempurl%
\url{https://www.openssl.org/}
\showURL{%
\tempurl}


\bibitem[\protect\citeauthoryear{??}{wol}{2017}]%
        {wolfSSL-SGX}
 \bibinfo{year}{2017}\natexlab{}.
\newblock \bibinfo{title}{{wolfSSL} with Intel$^{\circledR}$ {SGX}}.
\newblock
  \bibinfo{howpublished}{\url{https://www.wolfssl.com/wolfssl-with-intel-sgx/}}.
    (\bibinfo{year}{2017}).
\newblock


\bibitem[\protect\citeauthoryear{??}{SGX}{2018}]%
        {SGXWhitePaper18}
 \bibinfo{year}{2018}\natexlab{}.
\newblock \bibinfo{booktitle}{\emph{Performance Considerations for
  Intel$^{\circledR}$ Software Guard Extensions (Intel$^{\circledR}$ {SGX})
  Applications}}.
\newblock \bibinfo{type}{{T}echnical {R}eport}.
\newblock
\urldef\tempurl%
\url{https://software.intel.com/sites/default/files/managed/09/37/}
\showURL{%
\tempurl}


\bibitem[\protect\citeauthoryear{Arnautov, Trach, Gregor, Knauth, Martin,
  Priebe, Lind, Muthukumaran, O`Keeffe, Stillwell, et~al\mbox{.}}{Arnautov
  et~al\mbox{.}}{2016}]%
        {Scone16}
\bibfield{author}{\bibinfo{person}{Sergei Arnautov}, \bibinfo{person}{Bohdan
  Trach}, \bibinfo{person}{Franz Gregor}, \bibinfo{person}{Thomas Knauth},
  \bibinfo{person}{Andre Martin}, \bibinfo{person}{Christian Priebe},
  \bibinfo{person}{Joshua Lind}, \bibinfo{person}{Divya Muthukumaran},
  \bibinfo{person}{Daniel O`Keeffe}, \bibinfo{person}{Mark~L Stillwell},
  {et~al\mbox{.}}} \bibinfo{year}{2016}\natexlab{}.
\newblock \showarticletitle{SCONE: Secure linux containers with Intel SGX}. In
  \bibinfo{booktitle}{\emph{12th USENIX Symp. Operating Systems Design and
  Implementation}}.
\newblock


\bibitem[\protect\citeauthoryear{Aublin, Kelbert, O`Keeffe, Muthukumaran,
  Priebe, Lind, Krahn, Fetzer, Eyers, and Pietzuch}{Aublin
  et~al\mbox{.}}{2017}]%
        {aublin2017talos}
\bibfield{author}{\bibinfo{person}{Pierre-Louis Aublin},
  \bibinfo{person}{Florian Kelbert}, \bibinfo{person}{Dan O`Keeffe},
  \bibinfo{person}{Divya Muthukumaran}, \bibinfo{person}{Christian Priebe},
  \bibinfo{person}{Joshua Lind}, \bibinfo{person}{Robert Krahn},
  \bibinfo{person}{Christof Fetzer}, \bibinfo{person}{David Eyers}, {and}
  \bibinfo{person}{Peter Pietzuch}.} \bibinfo{year}{2017}\natexlab{}.
\newblock \bibinfo{booktitle}{\emph{{TaLoS}: Secure and transparent {TLS}
  termination inside {SGX} enclaves}}.
\newblock \bibinfo{type}{{T}echnical {R}eport}. \bibinfo{institution}{Technical
  Report 2017/5, Imperial College London}.
\newblock


\bibitem[\protect\citeauthoryear{Baumann, Peinado, and Hunt}{Baumann
  et~al\mbox{.}}{2014}]%
        {Haven14}
\bibfield{author}{\bibinfo{person}{Andrew Baumann}, \bibinfo{person}{Marcus
  Peinado}, {and} \bibinfo{person}{Galen Hunt}.}
  \bibinfo{year}{2014}\natexlab{}.
\newblock \showarticletitle{Shielding Applications from an Untrusted Cloud with
  Haven}. In \bibinfo{booktitle}{\emph{Proceedings of the 11th USENIX
  Conference on Operating Systems Design and Implementation}}
  \emph{(\bibinfo{series}{OSDI'14})}. \bibinfo{publisher}{USENIX Association},
  \bibinfo{address}{Berkeley, CA, USA}, \bibinfo{pages}{267--283}.
\newblock
\showISBNx{978-1-931971-16-4}
\urldef\tempurl%
\url{http://dl.acm.org/citation.cfm?id=2685048.2685070}
\showURL{%
\tempurl}


\bibitem[\protect\citeauthoryear{che Tsai, Porter, and Vij}{che Tsai
  et~al\mbox{.}}{2017}]%
        {Graphene17}
\bibfield{author}{\bibinfo{person}{Chia che Tsai}, \bibinfo{person}{Donald~E.
  Porter}, {and} \bibinfo{person}{Mona Vij}.} \bibinfo{year}{2017}\natexlab{}.
\newblock \showarticletitle{Graphene-SGX: A Practical Library {OS} for
  Unmodified Applications on {SGX}}. In \bibinfo{booktitle}{\emph{2017 {USENIX}
  Annual Technical Conference ({USENIX} {ATC} 17)}}. \bibinfo{pages}{645--658}.
\newblock


\bibitem[\protect\citeauthoryear{Chen, Wang, Jiang, Ding, Lu, Kim, Sahinalp,
  Shimizu, Burns, Wright, Png, Hibberd, Lloyd, Yang, Telenti, Bloss, Fox,
  Lauter, and Ohno-Machado}{Chen et~al\mbox{.}}{2017}]%
        {Princess17}
\bibfield{author}{\bibinfo{person}{Feng Chen}, \bibinfo{person}{Shuang Wang},
  \bibinfo{person}{Xiaoqian Jiang}, \bibinfo{person}{Sijie Ding},
  \bibinfo{person}{Yao Lu}, \bibinfo{person}{Jihoon Kim},
  \bibinfo{person}{S.~Cenk Sahinalp}, \bibinfo{person}{Chisato Shimizu},
  \bibinfo{person}{Jane~C. Burns}, \bibinfo{person}{Victoria~J. Wright},
  \bibinfo{person}{Eileen Png}, \bibinfo{person}{Martin~L. Hibberd},
  \bibinfo{person}{David~D. Lloyd}, \bibinfo{person}{Hai Yang},
  \bibinfo{person}{Amalio Telenti}, \bibinfo{person}{Cinnamon~S. Bloss},
  \bibinfo{person}{Dov Fox}, \bibinfo{person}{Kristin Lauter}, {and}
  \bibinfo{person}{Lucila Ohno-Machado}.} \bibinfo{year}{2017}\natexlab{}.
\newblock \showarticletitle{PRINCESS: Privacy-protecting Rare disease
  International Network Collaboration via Encryption through Software guard
  extensionS}.
\newblock \bibinfo{journal}{\emph{Bioinformatics}} \bibinfo{volume}{33},
  \bibinfo{number}{6} (\bibinfo{year}{2017}), \bibinfo{pages}{871}.
\newblock


\bibitem[\protect\citeauthoryear{Duan, Yuan, and Wang}{Duan
  et~al\mbox{.}}{2017}]%
        {LightBox17}
\bibfield{author}{\bibinfo{person}{Huayi Duan}, \bibinfo{person}{Xingliang
  Yuan}, {and} \bibinfo{person}{Cong Wang}.} \bibinfo{year}{2017}\natexlab{}.
\newblock \showarticletitle{LightBox: SGX-assisted Secure Network Functions at
  Near-native Speed}.
\newblock \bibinfo{journal}{\emph{CoRR}}  \bibinfo{volume}{abs/1706.06261}
  (\bibinfo{year}{2017}).
\newblock
\showeprint[arxiv]{1706.06261}
\urldef\tempurl%
\url{http://arxiv.org/abs/1706.06261}
\showURL{%
\tempurl}


\bibitem[\protect\citeauthoryear{Gueron}{Gueron}{2016}]%
        {Gueron16}
\bibfield{author}{\bibinfo{person}{Shay Gueron}.}
  \bibinfo{year}{2016}\natexlab{}.
\newblock \bibinfo{title}{A Memory Encryption Engine Suitable for General
  Purpose Processors}.
\newblock \bibinfo{howpublished}{Cryptology ePrint Archive, Report 2016/204}.
  (\bibinfo{year}{2016}).
\newblock
\newblock
\shownote{\url{https://eprint.iacr.org/2016/204}.}


\bibitem[\protect\citeauthoryear{Hoekstra, Lal, Pappachan, Phegade, and
  Del~Cuvillo}{Hoekstra et~al\mbox{.}}{2013}]%
        {HMR13SGX}
\bibfield{author}{\bibinfo{person}{Matthew Hoekstra}, \bibinfo{person}{Reshma
  Lal}, \bibinfo{person}{Pradeep Pappachan}, \bibinfo{person}{Vinay Phegade},
  {and} \bibinfo{person}{Juan Del~Cuvillo}.} \bibinfo{year}{2013}\natexlab{}.
\newblock \showarticletitle{Using Innovative Instructions to Create Trustworthy
  Software Solutions}. In \bibinfo{booktitle}{\emph{Proceedings of the 2nd
  international workshop on hardware and architectural support for security and
  privacy}}, Vol.~\bibinfo{volume}{13}.
\newblock


\bibitem[\protect\citeauthoryear{Knauth, Steiner, Chakrabarti, Lei, Xing, and
  Vij}{Knauth et~al\mbox{.}}{2018}]%
        {abs-1801-05863}
\bibfield{author}{\bibinfo{person}{Thomas Knauth}, \bibinfo{person}{Michael
  Steiner}, \bibinfo{person}{Somnath Chakrabarti}, \bibinfo{person}{Li Lei},
  \bibinfo{person}{Cedric Xing}, {and} \bibinfo{person}{Mona Vij}.}
  \bibinfo{year}{2018}\natexlab{}.
\newblock \showarticletitle{Integrating Remote Attestation with Transport Layer
  Security}.
\newblock \bibinfo{journal}{\emph{CoRR}}  \bibinfo{volume}{abs/1801.05863}
  (\bibinfo{year}{2018}).
\newblock
\showeprint[arxiv]{1801.05863}
\urldef\tempurl%
\url{http://arxiv.org/abs/1801.05863}
\showURL{%
\tempurl}


\bibitem[\protect\citeauthoryear{Ohrimenko, Schuster, Fournet, Mehta, Nowozin,
  Vaswani, and Costa}{Ohrimenko et~al\mbox{.}}{2016}]%
        {OSF+16ML}
\bibfield{author}{\bibinfo{person}{Olga Ohrimenko}, \bibinfo{person}{Felix
  Schuster}, \bibinfo{person}{Cedric Fournet}, \bibinfo{person}{Aastha Mehta},
  \bibinfo{person}{Sebastian Nowozin}, \bibinfo{person}{Kapil Vaswani}, {and}
  \bibinfo{person}{Manuel Costa}.} \bibinfo{year}{2016}\natexlab{}.
\newblock \showarticletitle{Oblivious Multi-Party Machine Learning on Trusted
  Processors}. In \bibinfo{booktitle}{\emph{25th USENIX Security Symposium
  (USENIX Security 16)}}. \bibinfo{address}{Austin, TX},
  \bibinfo{pages}{619--636}.
\newblock
\showISBNx{978-1-931971-32-4}


\bibitem[\protect\citeauthoryear{Orenbach, Lifshits, Minkin, and
  Silberstein}{Orenbach et~al\mbox{.}}{2017}]%
        {Eleos17}
\bibfield{author}{\bibinfo{person}{Meni Orenbach}, \bibinfo{person}{Pavel
  Lifshits}, \bibinfo{person}{Marina Minkin}, {and} \bibinfo{person}{Mark
  Silberstein}.} \bibinfo{year}{2017}\natexlab{}.
\newblock \showarticletitle{Eleos: ExitLess {OS} Services for {SGX} Enclaves}.
  In \bibinfo{booktitle}{\emph{Proceedings of EuroSys 2017}}.
  \bibinfo{pages}{238--253}.
\newblock


\bibitem[\protect\citeauthoryear{Schuster, Costa, Fournet, Gkantsidis, Peinado,
  Mainar{-}Ruiz, and Russinovich}{Schuster et~al\mbox{.}}{2015}]%
        {VC3}
\bibfield{author}{\bibinfo{person}{Felix Schuster}, \bibinfo{person}{Manuel
  Costa}, \bibinfo{person}{C{\'{e}}dric Fournet}, \bibinfo{person}{Christos
  Gkantsidis}, \bibinfo{person}{Marcus Peinado}, \bibinfo{person}{Gloria
  Mainar{-}Ruiz}, {and} \bibinfo{person}{Mark Russinovich}.}
  \bibinfo{year}{2015}\natexlab{}.
\newblock \showarticletitle{{VC3:} Trustworthy Data Analytics in the Cloud
  Using {SGX}}. In \bibinfo{booktitle}{\emph{2015 {IEEE} Symposium on Security
  and Privacy, {SP} 2015, San Jose, CA, USA, May 17-21, 2015}}.
  \bibinfo{pages}{38--54}.
\newblock


\bibitem[\protect\citeauthoryear{Wang, Wang, Bao, Wang, Wang, and Wu}{Wang
  et~al\mbox{.}}{2017}]%
        {WWBWWW17}
\bibfield{author}{\bibinfo{person}{Shuai Wang}, \bibinfo{person}{Wenhao Wang},
  \bibinfo{person}{Qinkun Bao}, \bibinfo{person}{Pei Wang},
  \bibinfo{person}{XiaoFeng Wang}, {and} \bibinfo{person}{Dinghao Wu}.}
  \bibinfo{year}{2017}\natexlab{}.
\newblock \showarticletitle{Binary Code Retrofitting and Hardening Using SGX}.
  In \bibinfo{booktitle}{\emph{Proceedings of the 2017 Workshop on Forming an
  Ecosystem Around Software Transformation}} \emph{(\bibinfo{series}{FEAST
  '17})}. \bibinfo{publisher}{ACM}, \bibinfo{pages}{43--49}.
\newblock


\bibitem[\protect\citeauthoryear{Weisse, Bertacco, and Austin}{Weisse
  et~al\mbox{.}}{2017}]%
        {Hotcalls17}
\bibfield{author}{\bibinfo{person}{Ofir Weisse}, \bibinfo{person}{Valeria
  Bertacco}, {and} \bibinfo{person}{Todd~M. Austin}.}
  \bibinfo{year}{2017}\natexlab{}.
\newblock \showarticletitle{Regaining Lost Cycles with HotCalls: {A} Fast
  Interface for {SGX} Secure Enclaves}. In
  \bibinfo{booktitle}{\emph{Proceedings of the 44th Annual International
  Symposium on Computer Architecture, {ISCA} 2017, Toronto, ON, Canada, June
  24-28, 2017}}. \bibinfo{pages}{81--93}.
\newblock
\urldef\tempurl%
\url{https://doi.org/10.1145/3079856.3080208}
\showDOI{\tempurl}


\bibitem[\protect\citeauthoryear{Zhang, Cecchetti, Croman, Juels, and
  Shi}{Zhang et~al\mbox{.}}{2016}]%
        {Towncrier16}
\bibfield{author}{\bibinfo{person}{Fan Zhang}, \bibinfo{person}{Ethan
  Cecchetti}, \bibinfo{person}{Kyle Croman}, \bibinfo{person}{Ari Juels}, {and}
  \bibinfo{person}{Elaine Shi}.} \bibinfo{year}{2016}\natexlab{}.
\newblock \showarticletitle{Town crier: An authenticated data feed for smart
  contracts}. In \bibinfo{booktitle}{\emph{Proceedings of the 2016 ACM SIGSAC
  Conference on Computer and Communications Security}}. ACM,
  \bibinfo{pages}{270--282}.
\newblock


\bibitem[\protect\citeauthoryear{Zheng, Dave, Beekman, Popa, Gonzalez, and
  Stoica}{Zheng et~al\mbox{.}}{2017}]%
        {Opaque17}
\bibfield{author}{\bibinfo{person}{Wenting Zheng}, \bibinfo{person}{Ankur
  Dave}, \bibinfo{person}{Jethro~G. Beekman}, \bibinfo{person}{Raluca~Ada
  Popa}, \bibinfo{person}{Joseph~E. Gonzalez}, {and} \bibinfo{person}{Ion
  Stoica}.} \bibinfo{year}{2017}\natexlab{}.
\newblock \showarticletitle{Opaque: An Oblivious and Encrypted Distributed
  Analytics Platform}. In \bibinfo{booktitle}{\emph{14th {USENIX} Symposium on
  Networked Systems Design and Implementation ({NSDI} 17)}}.
  \bibinfo{pages}{283--298}.
\newblock


\end{thebibliography}

\end{document}